\documentclass[twocolumn,showpacs,aps,pre,nofootinbib]{revtex4}
\usepackage{hyperref}
\usepackage{graphicx}
\usepackage{amsmath,amsfonts,amsthm}
\usepackage{color}

\usepackage{tikz}
\usetikzlibrary{arrows}
\usetikzlibrary{calc}
\usetikzlibrary{patterns}
\usetikzlibrary{scopes}

\newcommand{\<}{\big\langle}
\renewcommand{\>}{\big\rangle}

\begin{document}

\title{Anomalous critical behaviour in the polymer collapse transition  of three-dimensional lattice trails}

\author{Andrea Bedini} \email{abedini@ms.unimelb.edu.au}
\affiliation{Department of Mathematics and Statistics, The University
  of Melbourne, 3010, Australia}
\author{Aleksander L.\ Owczarek} \email{owczarek@unimelb.edu.au}
\affiliation{Department of Mathematics and Statistics, The University
  of Melbourne, 3010, Australia} 
\author{Thomas Prellberg} \email{t.prellberg@qmul.ac.uk}
\affiliation{School of Mathematical Sciences, Queen Mary University of
  London, Mile End Road, London, E1 4NS, United Kingdom}

\begin{abstract}
  Trails (bond-avoiding walks) provide an alternative lattice model of
  polymers to self-avoiding walks, and adding self-interaction at
  multiply visited sites gives a model of polymer collapse. Recently,
  a two-dimensional model (triangular lattice) where doubly and triply
  visited sites are given different weights was shown to display a
  rich phase diagram with first and second order collapse separated by
  a multi-critical point. A kinetic growth process of trails (KGT) was
  conjectured to map precisely to this multi-critical point. Two
  types of low temperature phases, globule and crystal-like, were encountered.

 Here, we investigate the collapse properties of a similar extended
 model of interacting lattice trails on the simple cubic lattice with
 separate weights for doubly and triply visited sites. Again we find
 first and second order collapse transitions dependent on the relative
 sizes of the doubly and triply visited energies. However we find no
 evidence of a low temperature crystal-like phase with only the
 globular phase in existence.

Intriguingly, when the ratio of the energies is precisely that which  separates the first
order from the second-order regions anomalous finite-sized scaling
appears. At the finite size location of the rounded transition clear
evidence exists for a first order transition that persists in the
thermodynamic limit. This location moves as the length increases, with
its limit apparently at the point that maps to a KGT. However, if one fixes the temperature to sit at
exactly this KGT point then only a critical point can be deduced from
the data. The resolution of this apparent contradiction lies in the
breaking of crossover scaling and the difference in the shift and
transition width (crossover) exponents.

 \end{abstract}



\maketitle
 
\section{Introduction}
\label{sec:intro}

The canonical lattice model of the configurations of a polymer in
solution has been the self-avoiding walk (SAW) where configurations
are those lattice paths that could be generated by a random walk on a
lattice that is not allowed to visit the same lattice site more than
once. Considered as a static equilibrium statistical mechanical
ensemble, self-avoiding walks display an excluded volume effect: they
are swollen in size if compared to the set of unrestricted random walks of the
same length. A common way to model intra-polymer interactions is to
assign an energy to each non-consecutive pair of monomers lying on
neighbouring lattice sites.  This prescription defines the interacting
self-avoiding walk (ISAW) model, which is the standard lattice model
of polymer collapse using self-avoiding walks. The collapse transition
in ISAW models, the so-called $\theta$-point of polymers, is a
second order phase transition that has been well studied. The standard
theory \cite{gennes1975a-a,stephen1975a-a,duplantier1982a-a} of the
collapse transition is based on the $n\rightarrow 0$ limit of the
magnetic tri-critical $\phi^4-\phi^6$ O($n$) field theory and related
Edwards model with two and three body forces
\cite{duplantier1986b-a,duplantier1987d-a}, which predicts an upper
critical dimension of three with subtle scaling behaviour in that
dimension. 

A physically equivalent way of obtaining the excluded volume in a
random walk model is to prevent the walk from visiting the same
\emph{bond} more than once. This weaker restriction leads to another
class of lattice paths called self-avoiding trails (SAT). The
interacting version of trails, customarily obtained by giving a weight
to multiply occupied sites, also presents a collapse
transition. The literature contains various definitions
\cite{meirovitch1990b-a,prellberg2001b-a,prellberg2001a-:a} of single
energy models of interacting self-avoiding trails (ISAT), all weighting multiply visited
sites in different ways (they differ in how sites visited more than
twice are assigned the energy of interaction). Regardless, theoretical
prediction \cite{shapir1984a-a} and the evidence
\cite{lim1988a-a,owczarek1995a-:a,prellberg1995b-:a,owczarek2006c-:a} suggests that the collapse transition
of the ISAT model is in a different universality class to that of ISAW,
although there is no clear understanding of why this is the case if
true.

On the other hand, a two-dimensional model (triangular
lattice) of an extended ISAT (eISAT) model, where doubly and triply
  visited sites are given \emph{different} energies, say
  $-\varepsilon_2$ and $-\varepsilon_3$ respectively, was recently \cite{doukas2010a-:a} shown to display a
  rich phase diagram with first and second order collapse separated by
  a multi-critical point. The occurrence of the type of transition
  depended on the ratio 
\begin{equation}
k= \frac{\varepsilon_3}{\varepsilon_2}
\end{equation}
of the energies given to multiply visited sites of different
degree. 

In conjunction with this study a stochastic process, known as kinetic
growth trails (KGT), was also considered, where configurations of trails
are produced by a growth process. The static configurations produced
are trails, but the trails of a fixed length are not all produced with
equal probability. The equivalent static model 
is an eISAT with a particular value of $k=k_{G}\simeq 4.15$ at one
particular temperature $T=T_G$. It was demonstrated \cite{doukas2010a-:a} that
this value of $k=k_G$ separates models where the collapse transition
is first order ($k > k_G$) from models where the collapse transition
is second order ($k<k_G$). It was shown that for $k < k_G$ the second
order transition was most likely in a single universality class, which
was the same as the one in the (two-dimensional) ISAW $\theta$-point collapse
transition. Moreover, the KGT process mapped exactly onto the
transition point separating the two lines of first and second order
transition. Defining the temperature of the transition in the
eISAT model to be $T_c(k)$, it was deduced that
\begin{equation}
T_c(k_G)= T_G\;.
\end{equation}
Also, importantly it was found that while the second order transition
encountered for small $k$ had a low temperature phase that was globular,
the low temperature phase at large $k$ was crystal-like, in that the
trail filled the lattice.

In this paper we investigate a counterpart to the
triangular lattice eISAT model on the simple cubic
lattice. One reason for choosing the simple cubic lattice  is that the
coordination number is the same as on the triangular lattice. 
This means that if one considers a
kinetic growth process KGT on the simple cubic lattice it will map to
precisely that same location in the phase diagram of the
three-dimensional model as the corresponding triangular lattice model
described above. Also the KGT on the simple cubic lattice has
previously been studied \cite{prellberg1995b-:a}: it demonstrates a
scaling behaviour consistent with it mapping to a critical point in
the phase diagram of eISAT. One might then speculate that the phase
diagram of the eISAT on the cubic lattice has the same structure as
on the triangular lattice. However, we find there are important
differences, including anomalous finite size scaling behaviour around the
KGT point, and no evidence of any crystal-like low temperature phase. 

The paper is set out as follows. We begin by recalling the theoretical
framework of the collapse transition. We then move onto defining the models
including the extended ISAT (eISAT) model on a general regular
lattice, specifically on the triangular and cubic lattices, the various
canonical ISAT models and the KGT process.  Before describing  our
results for the cubic lattice we summarise the findings on the
triangular lattice.

\section{Collapse Transition and Crossover Scaling}
\label{sec:collapse}

There are two related ways of describing the collapse. One is by
understanding the change in finite length $n$ scaling of key
quantities such as the radius of gyration, or alternatively end-to-end
distance, and partition function as the temperature is lowered past
the transition temperature $T_c$. The second is the associated
singular thermodynamic behaviour in the thermodynamic limit ($n\to\infty$) 
of the free energy per step and in the internal
energy and/or specific heat at that same temperature.

Let us first describe the finite size scaling change. As the
temperature is varied there is a collapse transition at some $T =
T_c$. For high temperatures ($T> T_c$) the excluded volume interaction
is the dominant effect, and the behaviour is universally the same as
the non-interacting SAW problem. The mean squared end-to-end distance
$R_n^2$  and partition function $Z_n$ are therefore expected to scale as
\begin{align}
  R_n^2 & \sim A\, n^{2\nu}\qquad \mbox{with } \quad \nu >1/2, \\\nonumber
  Z_n & \sim B\, \mu^n n^{\gamma - 1}.
\end{align}
respectively with estimates of $\nu$ and $\gamma$ to be $0.5874(2)$
\cite{prellberg2001b-a} and $1.156957(9)$ \cite{clisby2007a-a} in
three dimensions.

When fixed at the transition temperature $T=T_c$ for ISAW in three
dimensions, tri-critical field theory and the Edwards model
\cite{duplantier1986b-a,duplantier1987d-a} predict similar scaling
forms with $\nu=1/2$ and $\gamma=1$, though with additive logarithmic
corrections.

At low temperatures ($T > T_c$) it is accepted that the
partition function is dominated by configurations that are internally
dense. The partition function should then scale differently to that at
high temperature, since a collapsed polymer should have a well defined
surface (and associated surface energy). One expects \cite{owczarek1993e-:a} in three
dimensions asymptotics of the form
\begin{align}
  R_n^2 & \sim A\, n^{2/3},\\\nonumber
  Z_n & \sim B\, \mu^n \mu_s^{n^{2/3}} n^{\gamma-1}.
\end{align}
with $\mu_s < 1$. So the exponent $\nu=1/3$ and the fractal dimension
of the polymer becomes $3$.

This change in scaling behaviour is reflected in the thermodynamic
limit. In the thermodynamic limit there is expected to be a
singularity in the free energy, which can be seen in its second
derivative (the specific heat). Denoting the (intensive) finite length
specific heat \emph{per monomer} by $c_n(T)$, the thermodynamic limit
is given by the long length limit as
\begin{equation}
C(T) = \lim_{n\rightarrow\infty} c_n(T)\;.
\end{equation}
In general one expects that the singular part of the specific heat behaves as
\begin{equation}
C(T) \sim B |T_c - T|^{-\alpha}\;,
\end{equation}
where $\alpha<1$ for a second-order phase transition.
The singular part of the thermodynamic limit internal energy behaves as
\begin{equation}
U(T) \sim B |T_c -T|^{1-\alpha}\;,
\end{equation}
if the transition is second-order, and there is a jump in the internal
energy if the transition is first-order (an effective value of
$\alpha=1$).

Moreover one expects crossover scaling forms \cite{brak1993a-:a} to
apply around this temperature, so that
 \begin{equation}
  c_n(T) \sim n^{\alpha\phi}\,  \mathcal C([T-T_c]
  n^{\phi}),
  \label{crossover}
\end{equation}
with  $0<\phi < 1$ if the transition is second-order and 
  \begin{equation}
 c_n(T) \sim n \; {\cal C}([T - T_c]n)
\end{equation}
if the transition is first-order (that is, $\phi$ is effectively 1).  Assuming standard crossover theory
\cite{lawrie1984a-a} it was deduced in \cite{brak1993a-:a} that the
exponents $\alpha$ and $\phi$ are related via
\begin{equation}
\label{tricritical}
2-\alpha = \frac{1}{\phi}\;.
\end{equation}

Regardless of whether the full  crossover theory holds one can usually define
an exponent  $\psi$ from  the {\em shift} of the transition temperature at finite length
$T_{c,n}$ measured by, say, finding the position of a peak in the
specific heat as follows
\begin{equation}
  T_{c,n} - T_c \sim D\, n^{-\psi}.
\end{equation}
Another exponent can be related to the width $\Delta T_n$ of the transition.  We shall define this exponent $\phi$, via
\begin{equation}
  \Delta T_n \sim E\, n^{-\phi}.
\end{equation}
We use the same symbol $\phi$ because in a crossover theory the exponent will be the same $\phi$ as described in (\ref{crossover}).
For a first order transition one expects that both $\phi$ and $\psi$
are 1. For a second order transition in the standard scaling picture one expects the two exponents to be
equal, $\psi = \phi$, although we note a breaking down of the standard scaling
has been already observed in \cite{prellberg2002a-:a} in high dimensions.

For the ISAW model the tri-critical field theory expects the width (crossover)
exponent $\phi=1/2$ and shift exponent $\psi=1/2$ with logarithmic
corrections present in the scaling forms because the system is at the
upper critical dimension. The value of the specific heat exponent $\alpha=0$ is consistent with the scaling
relation (\ref{tricritical}), as 
a logarithmically divergent specific heat \cite{duplantier1986b-a,duplantier1987d-a} is predicted,
\begin{equation}
c_n(T_c) \sim C (\ln n)^{3/11} .
\end{equation}
The prediction \cite{duplantier1986b-a,duplantier1987d-a} for the shift
is
\begin{equation}
  T_{c,n} - T_c \sim D\,  n^{-1/2}(\ln n)^{-7/11}.
\end{equation}

On the other hand Shapir and Oono \cite{shapir1984a-a} have argued that the ISAT collapse
transition should be tri-critical in nature, as self-avoiding
walks. However they predict that ISAW and ISAT are in different
universality classes. Importantly, while the upper critical dimension
for ISAW is expected to be $d_u = 3$, the Shapir-Oono field theory
gives $d_u=4$ for ISAT.

\section{Models}
\label{sec:models}

\subsection{The general extended ISAT model (eISAT)}
\label{subsec:eISAT}

On a lattice with coordination number greater than 3\footnote{On a
  lattice with coordination number less than 4, a self-avoiding trail
  cannot visit the same site twice and therefore is just a
  self-avoiding walk} the self-interaction in the trail model can be
implemented in different ways, depending on how the weight associated
with contact site depends on how many times the site has been visited.
The canonical ISAT model, which has been defined differently by
different authors, fixes the energy associated with sites visited more
than twice based upon the energy of doubly visited sites. The eISAT is
a generalised model that allows for different (independent) energies
to be associated with multiply visited sites of different multiplicities.

Consider a regular lattice of coordination number $2q$ $(q \in \mathbb
N, q\geq2)$ and the configurations $\phi_n\in \Omega_n$ of trails of length $n$ (bonds)
starting from a fixed origin. Let $-\varepsilon_\ell$ be the energy
associated with lattice sites that have been visited $\ell$ times by
the trail.  Now let $m_\ell,\ \ell=1,\dotsc q$ be the number of
lattice sites that have been visited $\ell$ times by the trail. Note
that one always has $\sum_\ell\, \ell\, m_\ell = n + 1$\footnote{We will not count the initial occupation of the origin as a visit.}. Hence, to each of
these contact sites is associated an Boltzmann weight $\omega_\ell =
e^{\beta \varepsilon_\ell}$ with $\omega_1=1$, where $\beta=1/T$ is the 
inverse temperature in suitable units of the inverse Boltzmann constant. The partition
function is then given by:
\begin{equation}
  \label{eq:general-isat-z}
  Z_n(\omega_2, \dotsc, \omega_q) = \sum_{\phi_n \in \Omega_n}
  \omega_2^{m_2(\phi_n)} \dotsm \omega_q^{m_q(\phi_n)},
\end{equation}
and the probability distribution by
\begin{equation}
  \label{eq:pE}
  p_E(\phi_n; \omega_2, \dotsc, \omega_q) =
  \frac{\omega_2^{m_2(\phi_n)} \dotsm \omega_q^{m_q(\phi_n)}}
  {Z_n(\omega_2, \dotsc, \omega_q)} .
\end{equation}

We define a reduced finite-size free energy per step as
\begin{equation}
  \kappa_n = \frac{1}{n} \log Z_n,
\end{equation}
related to the usual free energy per step as $-\beta f_n = \kappa_n$.

The average of any quantity $Q$ over the ensemble of allowed paths
$\phi_n \in \Omega_n$ of length $n$ is given generically by
\begin{equation}
  \< Q \>_n = \frac{1}{Z_n} \sum_{\phi_n \in \Omega_n} Q(\phi_n)\ 
  \omega_2^{m_2(\phi_n)} \dotsm \omega_q^{m_q(\phi_n)}.
\end{equation}
The thermodynamic limit in this problem is given by the limit $n \to
\infty$, so that the thermodynamic free energy per step
$f_{\infty}$ is given by
\begin{equation}
  - \beta f_{\infty} = \kappa_{\infty} = \lim_{n \to
    \infty} \kappa_n .
\end{equation}
This quantity determines the asymptotic behaviour of the partition
function, i.e., $Z_n$ grows to leading order exponentially as
$\mu^{n}$ with $\mu = \exp \kappa_{\infty}$.

\subsection{Cubic and triangular lattice eISAT}
\label{subsec:kISAT}

Both the triangular and simple cubic lattices have coordination number
6. Therefore only two weights, $\omega_2$ and $\omega_3$, appear in the
formulae (\ref{eq:general-isat-z}) and (\ref{eq:pE}).

In order to explore the two-parameter space of the eISAT we define a
one-parameter family of models with weights defined by:
\begin{equation}
  \label{eq:kisat}
  (\omega_2,\, \omega_3) = (\omega, \omega^k)
  \qquad \text{($k$-eISAT)}
\end{equation}
for any positive real value of the parameter $k$, that is, the
energies obey
\begin{equation}
  \label{eq:kisatenergy}
   \varepsilon_3 = k \varepsilon_2
\end{equation}
in the $k$-eISAT model. We set the energy $\varepsilon_2=1$ for convenience from
now on.

In this parametrisation we define a reduced internal energy per step
and a reduced specific heat per step in the usual way via
\begin{align}
  \label{eq:def_u_c}
  u_n & = \partial \kappa_n / \partial \log \omega = \frac{\< m_2 +
    k\, m_3 \>_n}{n} \\
  c_n & = \partial u_n / \partial \log \omega = 
  \frac{\<(m_2 + k\, m_3)^2\>_n - \<m_2 + k\, m_3\>_n^2}{n}.
\end{align}

Let us define the collapse transition as occurring at 
\begin{equation}
\omega=\omega_c(k)
\end{equation}
so that we expect for fixed $k$ that
\begin{equation}
  c_n(\omega) \sim n^{\alpha\phi} \mathcal C((\omega-\omega_c) n^{\phi}),
\end{equation}
if crossover scaling occurs. More generally  the shift exponent is
defined by
\begin{equation}
\label{omega_psi}
  \omega_{c,n} - \omega_c \sim D\, n^{-\psi},
\end{equation}
where $\omega_{c,n}$ is the location of the peak of the specific heat,
and the width exponent by
\begin{equation}
\label{omega_phi}
  \Delta \omega_n \sim E\, n^{-\phi},
\end{equation}
where $\Delta\omega_n$ is the width of the half-height of the peak of
the specific heat. Both these exponents could depend on $k$.

\subsection{``Canonical'' ISAT models}
\label{subsec:canonical-isat-model}

The canonical model used by Doukas \textit{et al.}\
\cite{doukas2010a-:a} is one where every successive visit to a site adds an
energy $-\varepsilon$ to the total for that site so that a $q$-times
visited sites has energy $-(q-1)\varepsilon$. 
Therefore the canonical model is defined by the weight
parametrisation 
\begin{equation}
  \label{eq:canonical}
  (\omega_2, \omega_3) = (\omega, \omega^2)\;.
\end{equation}
The canonical model corresponds to the case $k = 2$ in our family of
interacting trails.  

A second ``canonical'' model used by Prellberg and Owczarek
\cite{prellberg1995b-:a, prellberg2001a-:a} has $k=3$
so that
\begin{equation}
  \label{eq:canonical2}
  (\omega_2, \omega_3) = (\omega, \omega^3)\;.
\end{equation}

In fact it seems that Meirovitch \textit{et al.}\ \cite{meirovitch1990b-a} used
$\omega_3=\omega_2$, that is $k=1$. Interestingly, all these models
$k=1,2,3$ have been seen \cite{doukas2010a-:a} to behave in the
same way on the triangular lattice and, as we shall see, seem to behave in
the same way on the simple cubic lattice (though the two- and three-dimensional models differ from each other in behaviour).

\subsection{Phase diagram of eISAT on the triangular lattice}
\label{subsec:phase-diagr-triang}

\begin{figure}[ht!]
  \includegraphics[width=0.9\columnwidth]{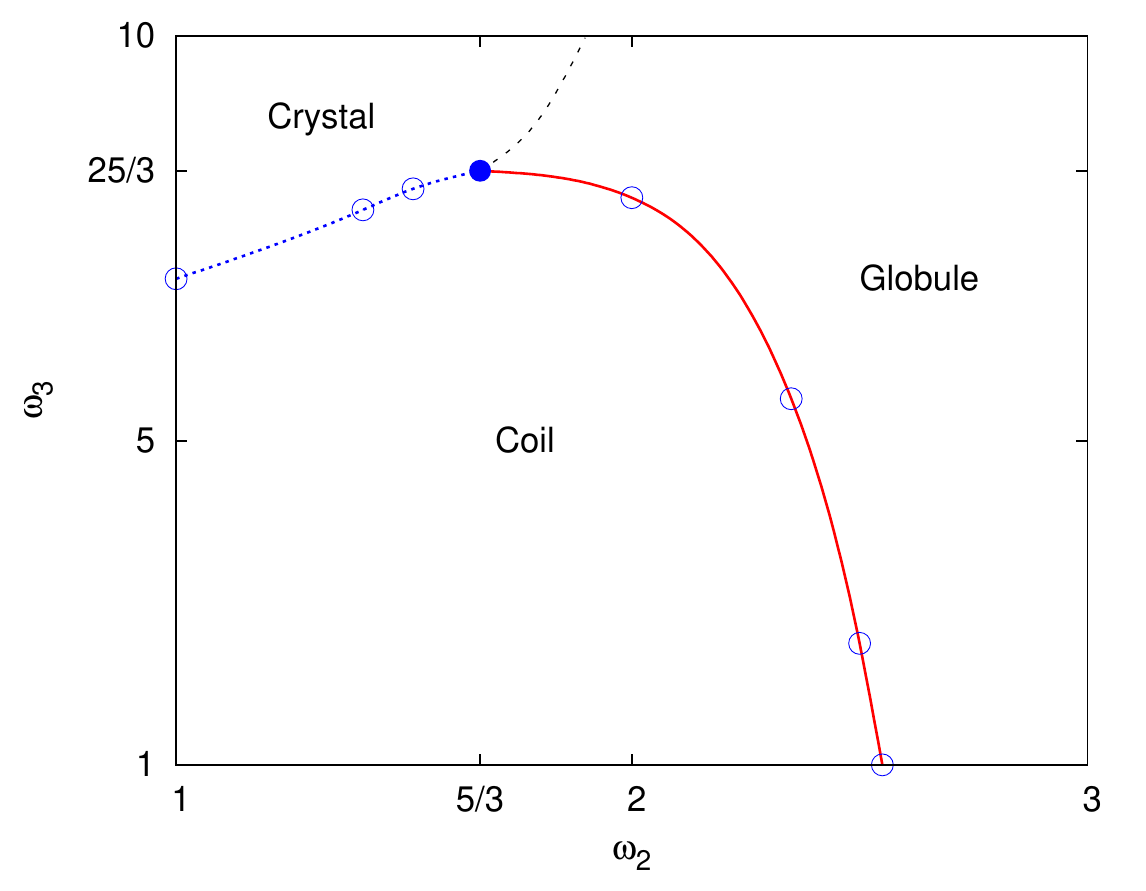}
  \caption{
  \label{fig:triangular-phase-diagram}
  Phase diagram of eISAT on the triangular lattice. The dashed transition
  line for small $\omega_2$ (large $k$) separating the coil from the crystal-like
  phases represents a first-order transition and
  ends at the solid circle which is the kinetic growth point
  (KGT). This KGT point is conjectured to be multi-critical. The solid
  line separating the coil from the globule represents the
  $\theta$-like second order transition found for small $k$. The line
  separating the globule from the crystal-like phase is conjectured to be
  second order but not $\theta$-like.
  }
\end{figure}

The study by Doukas \textit{et al.}\ \cite{doukas2010a-:a} on the triangular lattice identified
the kinetic growth point with a multi-critical collapse transition, being the  
meeting point of a swollen (coil), a collapsed and a crystal-like phase (see
Fig.~\ref{fig:triangular-phase-diagram}).

For small $\omega_2$ and $\omega_3$ the trails present the usual
swollen polymer phase where $\nu=3/4$ (in two dimensions). For large enough $\omega_2$,
regardless of $\omega_3$, there is a collapse phase, as occurs in the
ISAW model, and a transition between the swollen and collapsed globule
phases which seems to be $\theta$-like, with $\phi=3/7$ as expected in
two dimensions. On the other hand, for large
enough $\omega_3$ the ensemble is dominated by crystal-like
configurations that are space filling and internally contain only
triply-visited sites. Between the swollen phase and the crystal-like
phase the collapse transition is first-order. Separating this line of
first-order transitions from the line of $\theta$-like transitions is
a multi-critical point. This point is precisely the point
$(\omega^*_2, \omega^*_3)$ to which the kinetic growth process
of trails maps. The crystal-like and globule phase is separated by a
strong second order transition, much stronger than the $\theta$-collapse.

\section{The kinetic growth process (KGT)}
\label{sec:kinetic}

We now revisit the kinetic growth process \cite{lyklema1985a-a} of
trails (KGT).
Consider a regular lattice of coordination number $2q$ $(q \in \mathbb
N)$, and consider a stochastic process defined as follows: starting at
the origin, a lattice path is built up step-by-step by choosing
between available continuing steps from unoccupied lattice bonds with
equal probability. This dynamic process produces lattice paths that
are self-avoiding trails, moreover it is easy to show that, on a
coordination 6 lattice, a trail $\phi_n$ of length $n$ is generated
with probability
\begin{equation}
  p_G(\phi_n) = \frac{1}{6} \left(\frac{1}{5}\right)^{n-1}
  \left(\frac{5}{3}\right)^{m_2(\phi_n)}
  \left(\frac{25}{3}\right)^{m_3(\phi_n)}
  .
\end{equation}
This has to be compared with the probability distribution
(\ref{eq:pE}) of the equilibrium model with weights $(5/3, 25/3)$
\begin{equation}
  p_E\left(\phi_n; \frac{5}{3}, \frac{25}{3}\right) =
  \frac{1}  {Z_n(\frac{5}{3}, \frac{25}{3})}
  \left(\frac{5}{3}\right)^{m_2(\phi_n)}
  \left(\frac{25}{3}\right)^{m_3(\phi_n)}
  ,
\end{equation}
from which we can deduce
\begin{equation}
  p_G(\phi_n) \propto p_E\left(\phi_n; \frac{5}{3},
    \frac{25}{3}\right)
  .
\end{equation}
Note that the normalisation is different since the sum over all walks
of fixed length gives the probability of walks being still open in the
case of the growth process, and unity in the case of the equilibrium
model.

We shall refer to the weight choice to which the KGT maps to as the
Kinetic Growth point in the $(\omega_2, \omega_3)$ plane:
\begin{equation}
  \label{eq:kinetic_growth_point}
  (\omega^*_2, \omega^*_3) = \left(\frac53, \frac{25}3 \right)
  \qquad \text{(KGT)}
  .
\end{equation}
The Kinetic Growth point does not correspond to any point in any of
the canonical parametrisations described above
(\ref{subsec:canonical-isat-model}) but it belongs to our family of
interacting trails (\ref{eq:kisat}) with $k$ equal to
\begin{equation}
  k_G \equiv \frac{\log 25/3}{\log 5/3} \simeq 4.15 \ldots\; .
\end{equation}
The KGT point has
\begin{equation}
  T_G \equiv \frac{1}{\log 5/3} \simeq 1.957\dots \; .
\end{equation}
In order to match the definition in (\ref{eq:def_u_c}) at the KGT
point,
\begin{align}
  \label{eq:coorspond_u_c}
  {u}_n(\omega^*_2, \omega^*_3) & = {u}_n^* \\
  {c}_n(\omega^*_2, \omega^*_3) & = {c}_n^*
  ,
\end{align}
we define analogues of the internal energy and specific heat
by
\begin{align}
  \label{eq:defkinetic_u_c}
  {u}_n^* & = \frac{\< m_2 + k_G\, m_3 \>_n}{n} \\
  {c}_n^* & = \frac{\<(m_2 + k_G\, m_3)^2\>_n - \<m_2 + k_G\,
    m_3\>_n^2}{n}.
\end{align}

Kinetic Growth trails have been subject of various studies both in low
and high dimensionality and on a variety of lattices (see, for
example, \cite{owczarek1995a-:a,prellberg1995b-:a,doukas2010a-:a}).

Simulations on the simple cubic lattice show a divergent specific heat
with a logarithmic divergence different from the one
predicted by the Edwards model.

In particular it was seen that
\begin{equation}
  {c}_n^* \sim C^* (\log n)^{\zeta}
\end{equation}
with
\begin{equation}
  \zeta = 1.0 \pm 0.5 .
\end{equation}
whereas the Edwards model has $\zeta=3/11$.

\section{Results for the cubic lattice eISAT}
\label{sec:results}

\subsection{The KGT point}
\label{subsec:at-kinetic-growth}

\begin{figure}[ht!]
  \includegraphics[width=\columnwidth]{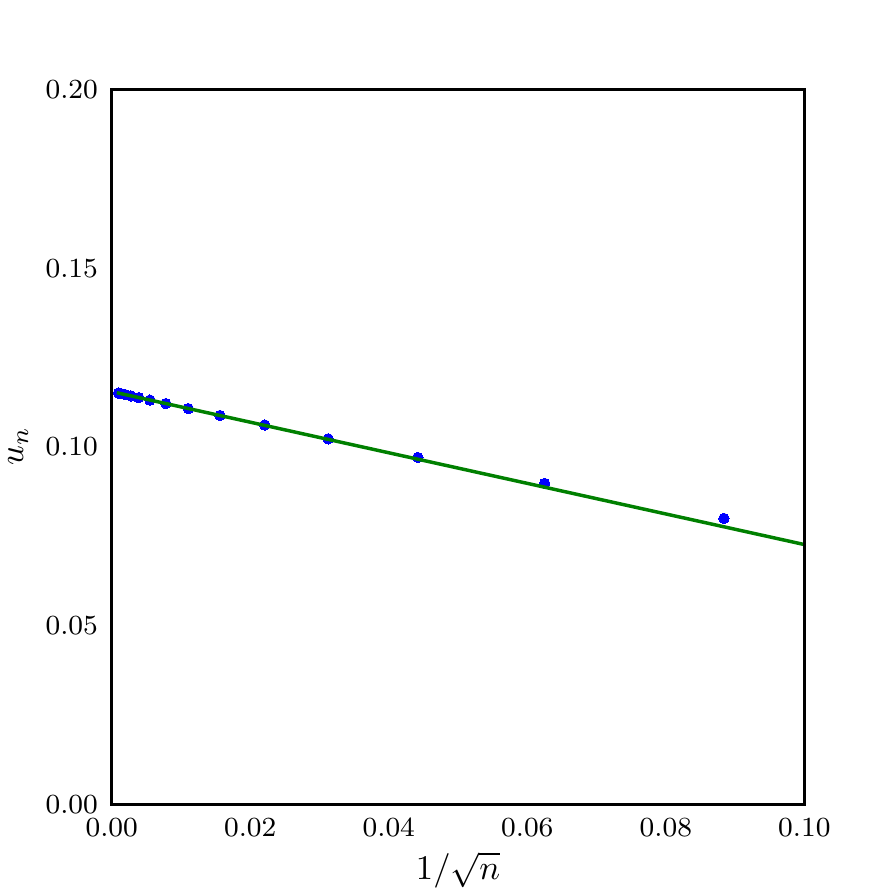}

  \caption{Plot of the energy $u_n$ of kinetic growth trails on the 
       simple cubic lattice against $1/\sqrt{n}$, along with a line of best fit.
    }
  \label{fig:kgw:un}
\end{figure}

\begin{figure}[ht!]
  \includegraphics[width=\columnwidth]{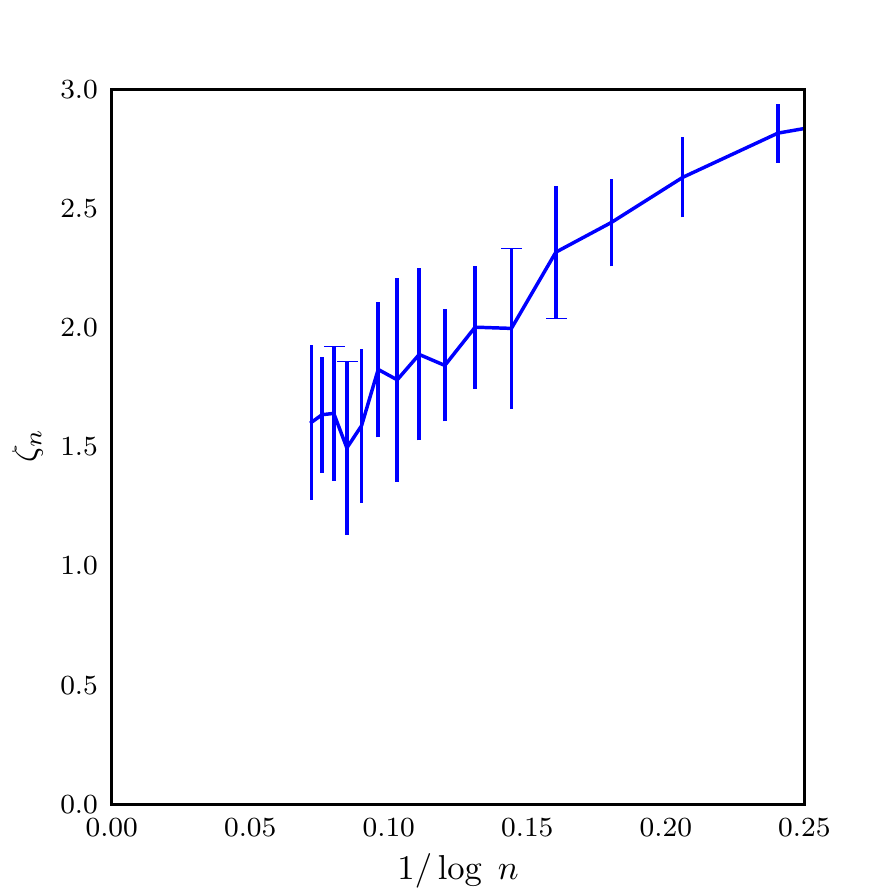}
  \caption{Estimating the exponent $\zeta$ in Eqn. (\ref{zetaeqn}) for KGT:
    Local slopes $\zeta_n$ of $\log c_n$ versus $\log \log n$ plotted
    against $1/ \log n$. Error bars are obtained by a simple re-binning
    procedure.}
\label{fig:kgw_zeta_estimate}
\end{figure}

We begin our investigation by revisiting the KGT point on the simple
cubic lattice directly by simulating the eISAT at that point. That is, we have simulated the KGT point of the eISAT
model at
\begin{equation}
(\omega_2, \omega_3) =(\omega^*_2, \omega^*_3) \equiv \left(\frac53, \frac{25}3 \right)\;.
\end{equation}

We simulated $10^5$ realisations of the kinetic growth process
collecting $8.7 \times 10^4$ samples of length $2^{20}$. We then
computed the energy $u_n$ and the specific heat $c_n$. The results
confirm what already was reported in
\cite{prellberg1995b-:a}: $u_n$ behaves as
\begin{equation}
u_n \sim u_{\infty} - \frac{u}{\sqrt{n}}\;,
\end{equation}
as indicated in Fig. \ref{fig:kgw:un},
and the specific heat $c_n$ diverges as a power of $\log n$
\begin{equation}
c_n \sim c\, (\log n)^{\zeta}.
\label{zetaeqn}
\end{equation}
We calculated the local slopes $\zeta_n$ of $\log c_n$ versus $\log
\log n$, which are plotted versus $1/\log n$ in
Fig.~\ref{fig:kgw_zeta_estimate}. We estimate $\zeta \simeq 1.0 \pm
0.5$ as reported in \cite{prellberg1995b-:a}.

\subsection{The $k_G$-eISAT model}
\label{subsec:kg-eISAT}

We have first simulated the $k$-eISAT model (\ref{eq:kisat}) with $k =
k_G$ (the correspondence with KGT occurs at $\omega = 5/3$) using 
the flatPERM algorithm
\cite{prellberg2004a-a,janse2009a-a}. We ran $S = 10^5$ iterations of
generating about $S_n \simeq 9 \cdot 10^9$
samples at length $10^3$. Following \cite{prellberg2004a-a}, we also
measured the number of samples adjusted by the number of their
independent growth steps, obtaining $S_n^{eff} \simeq 4 \cdot 10^7$
``effective'' samples.

FlatPERM outputs an estimate $W_{n,m}$
of the total weight of the walks of length $n$ and fixed value of
$m$. From the total weight one can access physical quantities over a
broad range of temperatures through a simple weighted average, e.g.
\begin{align}
  u_n(\omega) &= \frac{\sum_m m\, W_{n,m}}{\sum_m W_{n,m}}\quad\text{and}
\\
  c_n(\omega) &= \frac{\sum_m m^2\, W_{n,m}}{\sum_m W_{n,m}} - \left(
    \frac{\sum_m m\, W_{n,m}}{\sum_m W_{n,m}}\right)^2 \;.
\end{align}
The analysis of the scaling of the specific heat peak is done by
calculating the location of the peak of the specific heat $\omega_{c,n}$ 
and thereby evaluating $c_n^{peak} = c_n(\omega_{c,n})$. 

\begin{figure}[ht!]
  \includegraphics[width=\columnwidth]{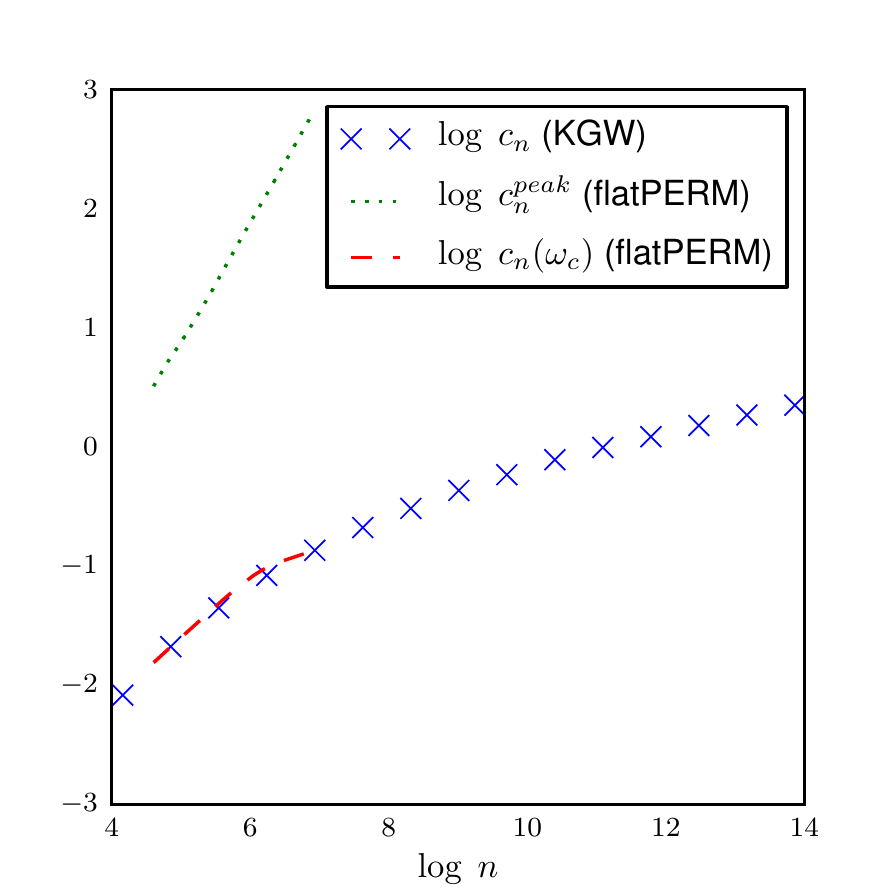}
  \caption{Comparison of the growth of the specific heat at the kinetic
growth point (lower dashes and crosses) with the growth of the peak of the specific
heat (upper dots) for eISAT with $k=k_G$. One clearly sees different growth rates in the upper and lower
curves, in particular the peak of the specific heat (upper dots) grows linearly on this logarithmic scale.}
  \label{fig:c_kgw_vs_flatperm}
\end{figure}

At the kinetic growth point, we confirm the slow growth of the specific heat 
seen in the previous section (c.f. Eqn. (\ref{zetaeqn})). The KGT simulations
and flatPERM simulations evaluated at $\omega_c$ coincide as shown in Fig. \ref{fig:c_kgw_vs_flatperm}.

Somewhat unexpectedly, the height of the peak of the specific heat, $c^{peak}_n$,
shows a different behaviour, diverging as a power of $n$, as illustrated by the top curve in
Fig.~\ref{fig:c_kgw_vs_flatperm}. In fact the divergence is linear, which 
indicates a first order transition. The nature of the finite-size
transition is therefore very different from the one inferred from the
kinetic growth process.

\begin{figure}[ht!]
  \includegraphics[width=\columnwidth]{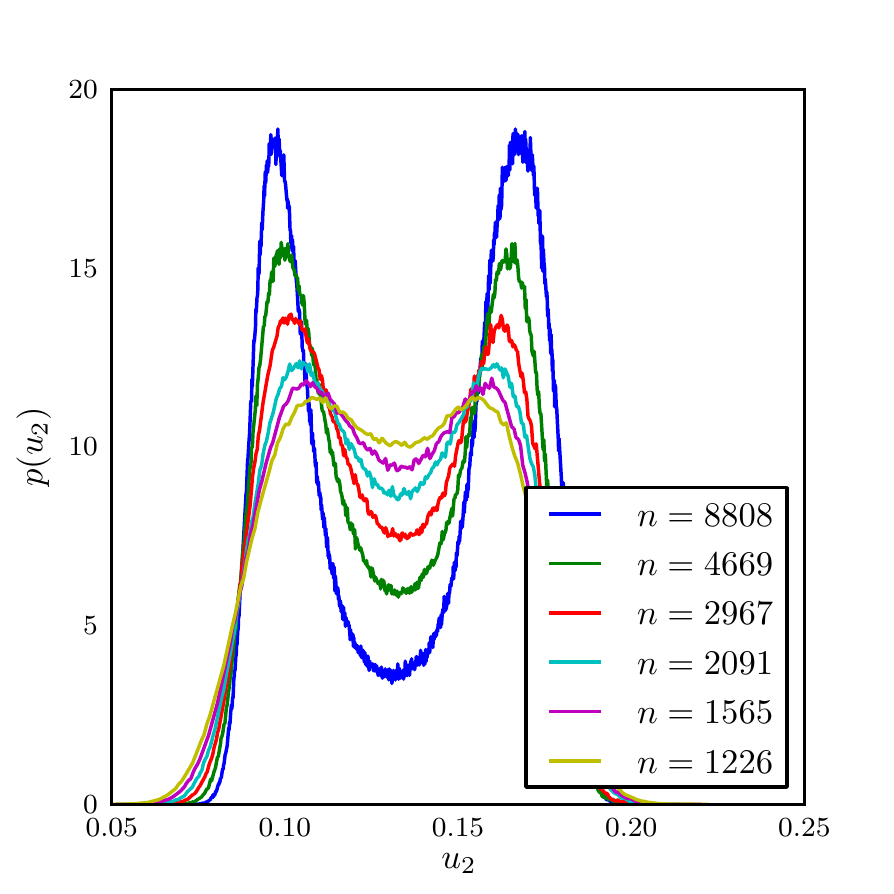}
  \caption{Normalised energy density of states near the kinetic-growth point for eISAT with $k=k_G$. The values of $\omega$
  have been chosen such that both peaks have equal height. One clearly sees the build-up of a first-order
  transition with a well defined gap between the two peaks. This will result in a latent heat in the thermodynamic
  limit.}
  \label{fig:kG_double_peaks}
\end{figure}

We have therefore simulated this model at longer lengths ($10^4$) using a thermal
implementation of flatPERM at several fixed
temperatures around the expected location of the transition. We
have found that the energy distribution shows indeed a double peak
(see Fig.~\ref{fig:kG_double_peaks}) with the two peaks getting
more and more definite as the length scale increases. 

We infer from this that a simulation at $\omega_c$ does not ``see'' the second
peak and hence only shows the observed much weaker divergence of the specific heat.

This is consistent with a scenario in which the shift and width of the transition
scale with different exponents as in Eqns. (\ref{omega_psi}) and (\ref{omega_phi}).

\begin{figure}[ht!]
    \includegraphics[width=\columnwidth]{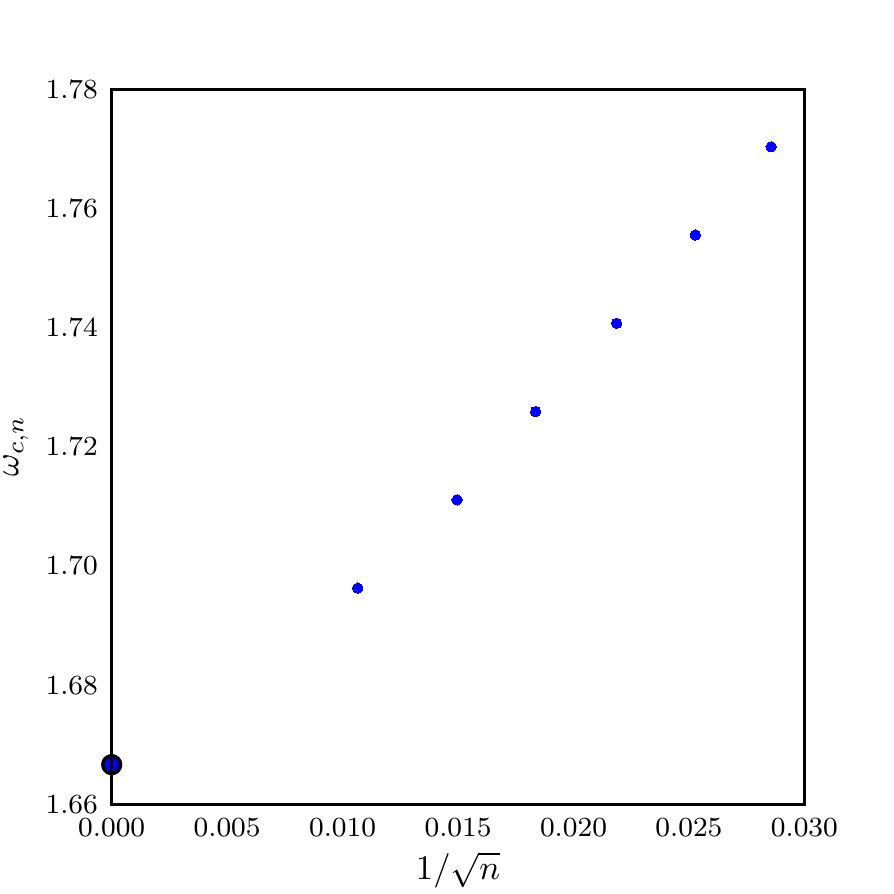}
  \caption{The peak position of the specific heat $\omega_{c,n}$ versus $1/\sqrt{n}$ for eISAT with $k=k_G$. The location of the kinetic growth point
at $\omega=5/3$ is marked with a large dot on the vertical axis. One can see that the peak position of the specific heat approaches this point in the thermodynamic limit.}
   \label{fig:kg_shift_scaling}
\end{figure}

\begin{figure}[ht!]
    \includegraphics[width=\columnwidth]{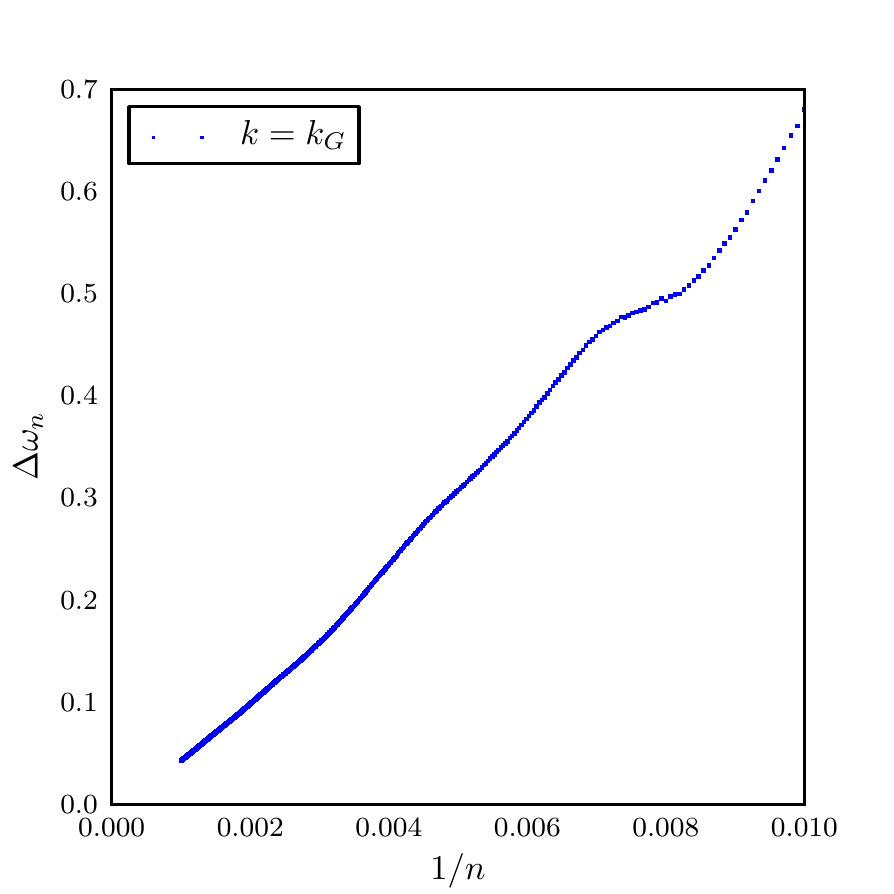} 
  \caption{The half width the specific heat $\Delta\omega_n$ versus $1/n$ for eISAT with $k=k_G$.}
 \label{fig:kg_peak_width_scaling}
\end{figure}

Fig. \ref{fig:kg_shift_scaling} indicates that the peak position of the specific heat
converges to the kinetic growth point $\omega_c = 5/3$ with an expected inverse square-root like scaling,
while Fig. \ref{fig:kg_peak_width_scaling} shows that the width of the transition decreases much
faster with an inverse linear scale.

We thus conclude that
\begin{equation}
  \omega_{c,n} - \omega_c \sim D\, n^{-1/2}\;,
\end{equation}
where $\omega_{c,n}$ is the location of the peak of the specific heat, 
and
\begin{equation}
  \Delta \omega_n \sim E\, n^{-1}\;,
\end{equation}
where $\Delta \omega_n$ is the half-width (or, more precisely, width of the half-height) of the
specific heat peak.

\begin{figure}[ht!]
  {\includegraphics[width=\columnwidth]{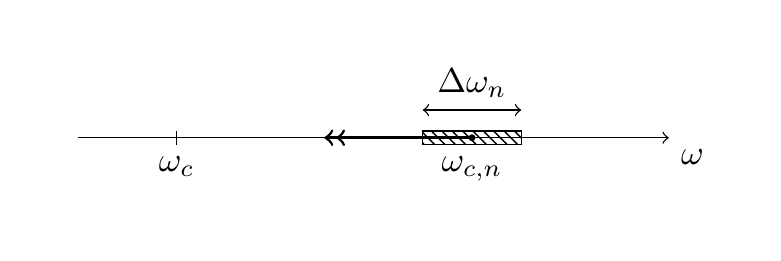}}
  \caption{The observed scaling at $k = k_G$ is such that
    $(\omega_{c,n} - \omega_{c}) \gg \Delta\omega_n$.}
  \label{fig:scaling}
\end{figure}

Since the width decays much faster than the shift, 
one cannot see the
first-order nature of the true thermodynamic transition 
in a kinetic growth
simulation which is fixed at the transition point;
it lies outside the crossover region. 
This is graphically indicated in Fig.~\ref{fig:scaling}.

\subsection{$k$-eISAT with $k>k_G$}
\label{subsec:large-k}

We next discuss the case of $k>k_G$, where on the triangular lattice a
first-order transition was found.

We simulated $k=5$, $k=7$,
and a model where only triply visited sites are weighted (this can be seen
as the limit of letting $k\to\infty$).
We simulated trails with length up to $10^3$, collecting at that length
$S_n \simeq 7 \cdot 10^8$, $10^9$ and $2 \cdot 10^8$ samples
corresponding respectively to $S_n^{eff} \simeq 2 \cdot 10^7$, $3
\cdot 10^7$ and $4 \cdot 10^6$ effective samples.

\begin{figure}[ht!]
  {\includegraphics[width=\columnwidth]{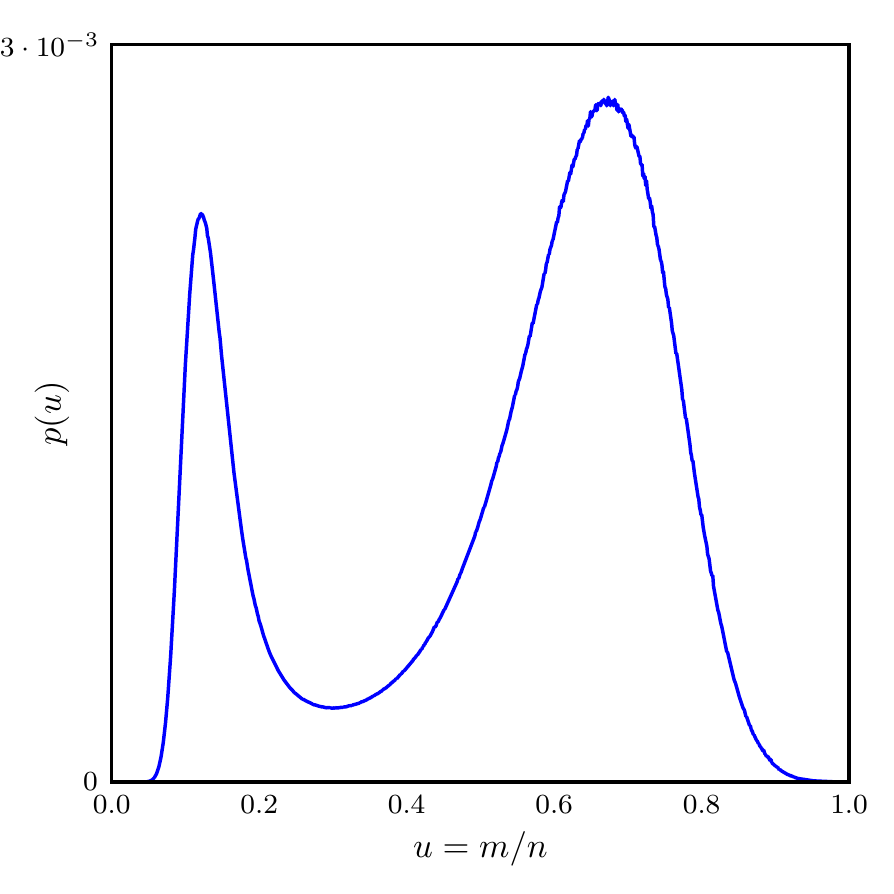}}

  {\includegraphics[width=\columnwidth]{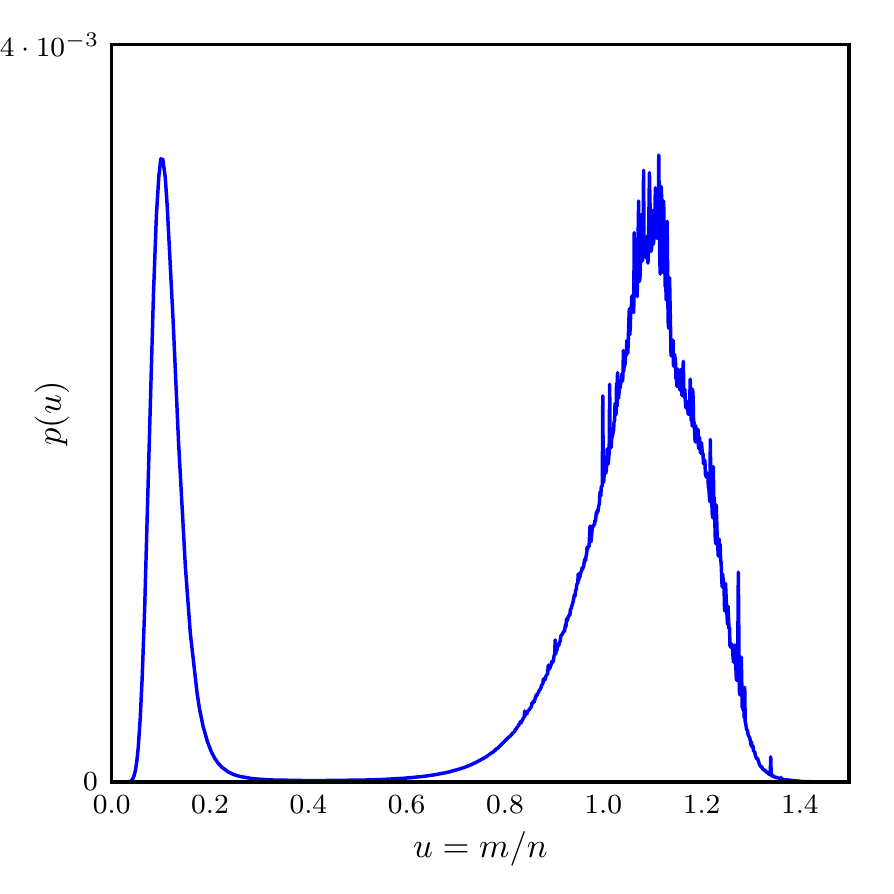}}

  \caption{Normalised energy density of states ($k > k_G$) for $k=5$ and $7$, at the respective values of $\omega=1.65$ and 
   $1.44$, i.e. near the peak of the specific heat curves.}
  \label{fig:k5_k7_double_peak}
\end{figure}

The energy distribution displays a clear double peaked form
(Fig. \ref{fig:k5_k7_double_peak}) which becomes more pronounced as $k$ 
increases.
Moreover, it sharpens as $n$ increases, which is clear evidence of a
first-order phase transition. We find that due to the initial build-up
of the bi-modality the specific heat actually seems to increase faster than
linearly. The scaling of the shift and width of the transition here is no
longer consistent with the scenario found at $k=k_G$. We now find evidence
for $1/n$ scaling of the shift of the transition, rather than the $1/\sqrt n$ scaling
found at $k=k_G$.

\subsection{$k$-eISAT with $k<k_G$}
\label{subsec:small-k}

Finally we focus on $k$-eISAT for $k<k_G$. We simulated the $k$-eISAT model 
with $k=0,1,2,3$, collecting for each
simulation order of $S_n \simeq 10^8$ samples (corresponding to
$S_e^{eff} \simeq 10^6$) at the maximum length $N = 10^3$. 

On the triangular lattice, for $k < k_G$ the collapse transition is 
$\theta$-like, and hence second order. A corresponding conjecture for the
cubic lattice is the presence of a weak second-order transition  with  a
logarithmically divergent specific heat $c_n \simeq (\ln\, n)^{\zeta}$.

\begin{figure}[ht!]
  \includegraphics[width=\columnwidth]{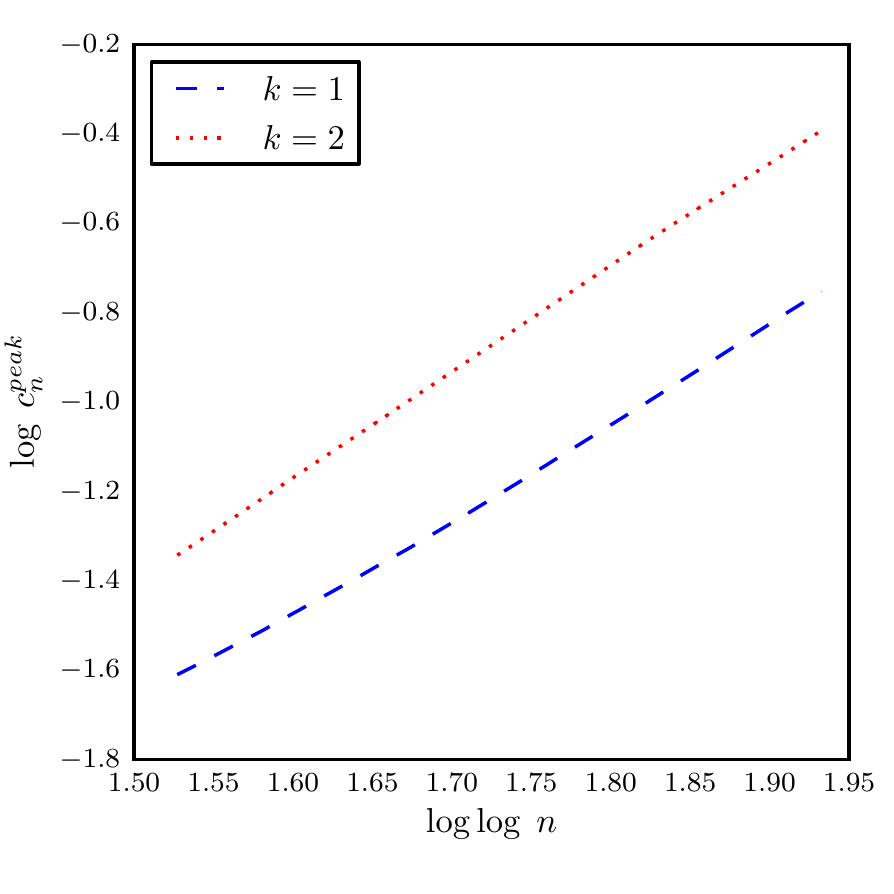}
  \caption{Specific heat peak scaling for $k=1$ and $k=2$: the specific heat
  diverges logarithmically with an exponent close to $\zeta=2.25$.}
  \label{fig:k1_k2_specific_heat_scaling}
\end{figure}

The results for $k = 1$ and $k = 2$, which we report in
Fig.~\ref{fig:k1_k2_specific_heat_scaling}, are consistent with
this prediction. However, we note that our estimated value $\zeta = 2.25 \pm 0.25$
is outside the prediction $\zeta=3/11$ for the $\theta$-point.

While models with $k = 0$ and $k = 3$ show very strong scaling
corrections which make the analysis inconclusive there, we expect the
second-order phase transition scenario to extend to these values, and indeed
to the whole range of values $0\leq k<k_G$.

\subsection{Low temperatures}
\label{subsec:crystal}

Motivated by the results on the triangular lattice, we now investigate
the possible presence of a crystal-like phase in three dimensions. 

\begin{figure}[ht!]
  {\includegraphics[width=\columnwidth]{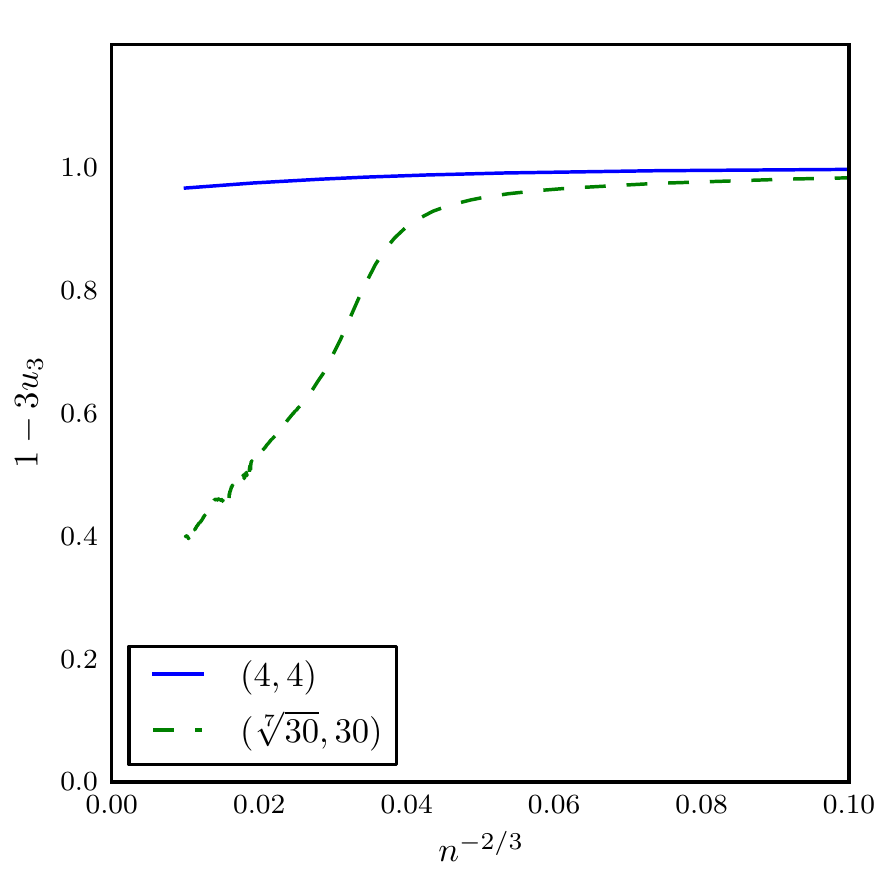}}
  \caption{$1-3u_3$, the proportion of steps not involved with
    triply-visited sites per unit length, plotted against $n^{-2/3}$,
    for $(\omega_2,\omega_3)=(4,4)$ and $(\sqrt[7]{30},30)$,
    respectively.}
  \label{fig:k2_k5_k7_triple_contacts}
\end{figure}

In a crystal-like phase, in which the trail asymptotically fills the lattice, 
the quantity $1-3 u_3$, i.e. the proportion of steps that are not involved with
triply-visited sites per unit length, should tend to zero as $n\to\infty$. Based 
on this criterion,
the investigation on the triangular lattice identified two different regions of the 
collapsed phase: one in which $1-3u_3$ tended to zero, and one in which it did not.
The one in which it did was associated with the first-order transition for $k>k_G$.

Following the analysis in \cite{doukas2010a-:a}, we show in
Fig.~\ref{fig:k2_k5_k7_triple_contacts} the quantity $1-3 u_3$
against $n^{-2/3}$ at two points in the parameter region of the low-temperature phase. 
This is the expected order of finite-size correction 
due to the presence of a surface in a compact low-temperature cluster. The parameters
$(\omega_2,\omega_3)$ chosen are representative of regions in the $\omega_2,\omega_3$--plane 
for which we would expect to observe a significant difference. While there is a substantial 
difference in the asymptotic value of $1-3u_3$ in these regions, it is clear that $1-3u_3$ does
not tend to zero in either.

We have also looked for a signature of a transition between low
temperature phases and were unable to find any emerging transition,
unlike for the triangular lattice (see
Fig.~\ref{fig:finite-size-phase-diagram} below). Of course we cannot exclude that such
a transition may become apparent at longer trail lengths.

\section{Phase diagram}
\label{sec:phasediagram}

To investigate further the full two-dimensional phase diagram, we ran two
more sets of simulations. First we ran a two-parameter flatPERM
simulation of the general eISAT model. A simulation of this type is
limited by both time and memory requirement but we have been
able to collect $S_n^{eff} \simeq 10^8$ effective samples at length $N
= 512$.

\begin{figure}[ht!]
  \includegraphics[width=\columnwidth]{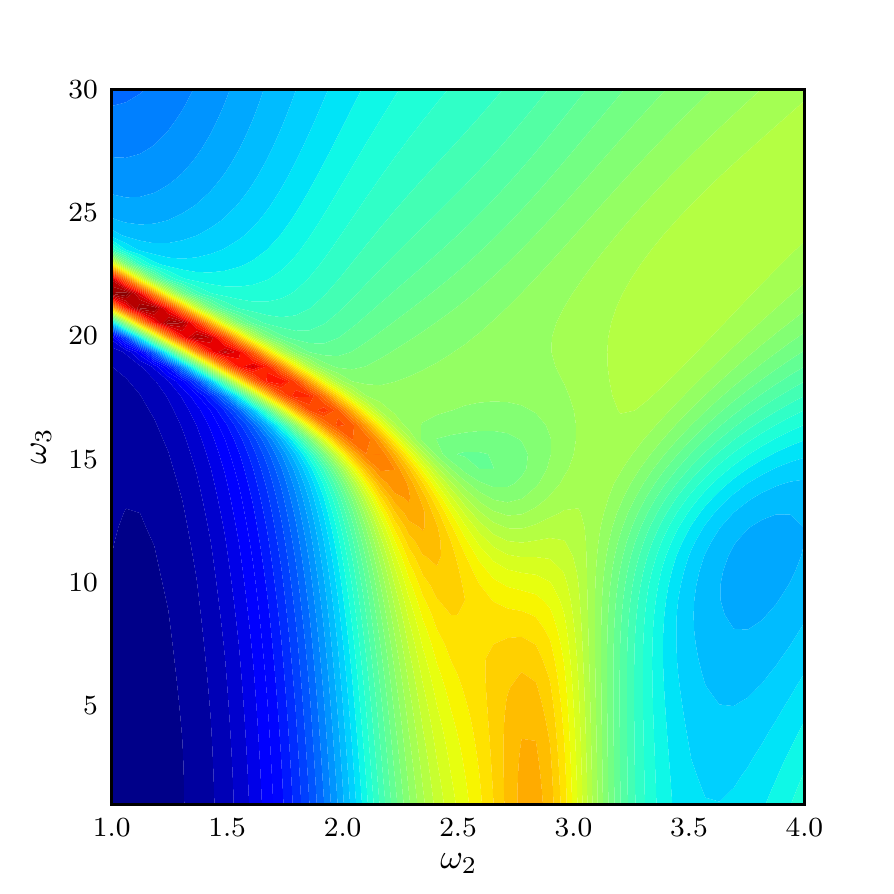}
  \caption{Density plot of the logarithm of the largest
    eigenvalue $\lambda_{max}$ of the matrix of second derivatives of
    the free energy with respect to $\omega_2$ and $\omega_3$ at
    length $n = 512$.}
\label{fig:eigenvalue_plot}
\end{figure}

\begin{figure}[ht!]
  {\includegraphics[width=\columnwidth]{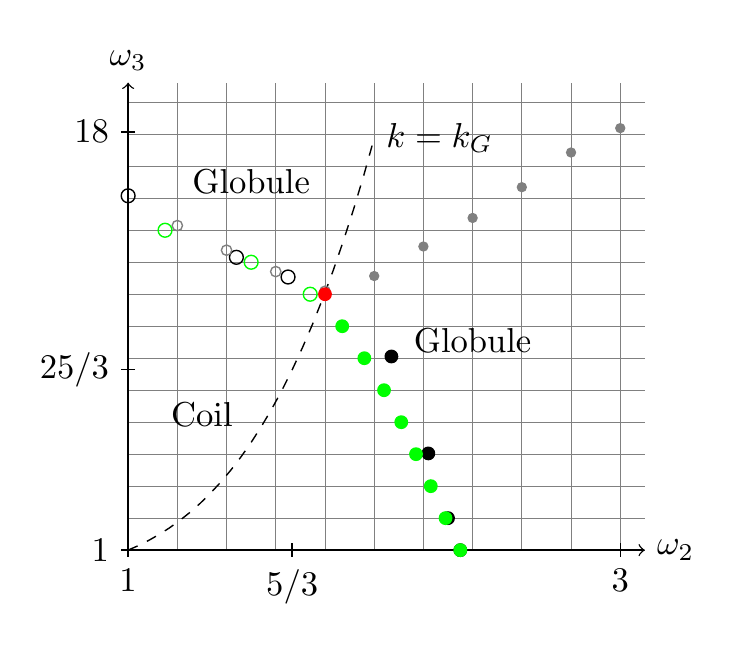}}
  \caption{Schematic of the observed phase diagram
    at length $n = 1000$. Dots indicate the location of peaks of fluctuations.
    Empty and filled dots indicate putative first and second order 
    phase transitions, respectively, whereas small grey dots indicate crossover.
    The dot on the $k=k-G$-curve corresponds to the special transition described in
    Sec.~\ref{subsec:at-kinetic-growth}. }
\label{fig:finite-size-phase-diagram}
\end{figure}

A density plot of the maximum fluctuations, calculated from the
eigenvalues of the matrix of second derivatives of the free energy
is shown in Fig.~\ref{fig:eigenvalue_plot}. Our
inference for the finite size phase diagram is shown in
Fig.~\ref{fig:finite-size-phase-diagram}.

Then we investigated the phase diagram along vertical and horizontal
slices, that is to say with $\omega_2$ and $\omega_3$ fixed, for $n$ up to 1000. Observed
phase transition points are included in
Fig.~\ref{fig:finite-size-phase-diagram}. 
In the collapsed region we find evidence for a wide crossover region, but no
evidence for an actual transition between two distinct phases.

\begin{figure}[ht!]
  {\includegraphics[width=\columnwidth]{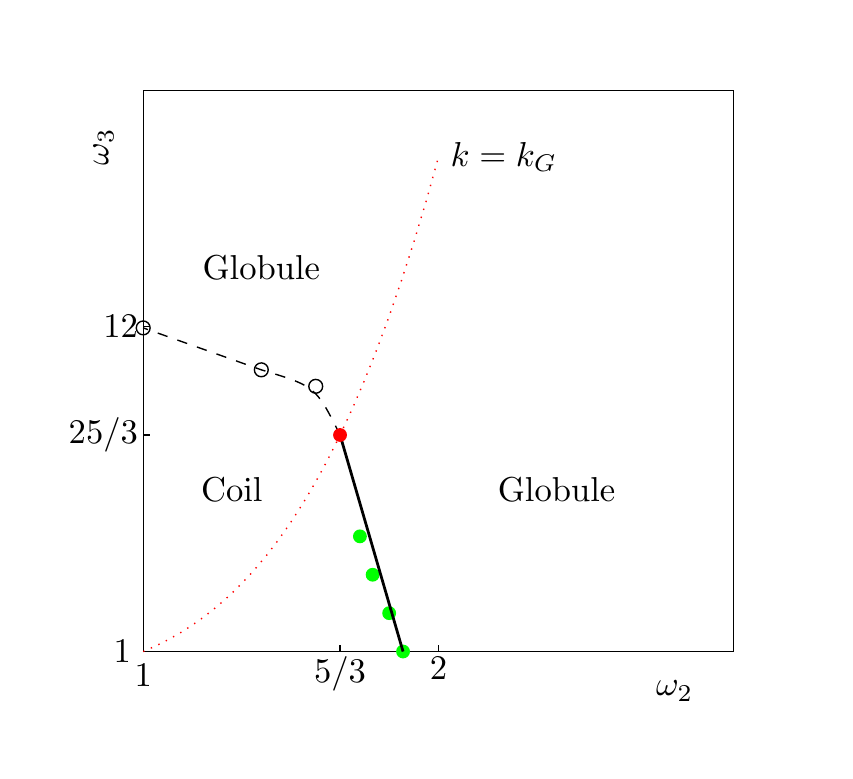}}
  \caption{Conjectured phase diagram for cubic lattice.  Solid
    lines indicate phase transitions of the second order while the
    dashed indicates a first-order phase transition. 
    The red
    dotted line only indicates the $k$-eISAT model which passes
    through the KGT point. Empty and filled dots indicate an
    estimated location of the phase transitions coloured as in
    Fig.~\ref{fig:finite-size-phase-diagram}.}
  \label{fig:conjecture-pd}
\end{figure}

We have provided numerical evidence of that the two different types of
transitions lead to a single collapsed globule-like phase at low
temperatures. Our conjecture for the thermodynamic phase diagram is
found in Fig.~\ref{fig:conjecture-pd}.\\

\section{Conclusions}
\label{sec:conclusions}

We have investigated the collapse properties of an extended family of
interacting self-avoiding trails in three dimensions on the simple
cubic lattice where doubly and triply visited sites are given weights
$\omega_2=e^{\varepsilon_2/T}$ and $\omega_3=e^{\varepsilon_3/T}$.

We have explored the general eISAT model by considering a family of
models satisfying $\omega_3 = \omega_2^k$ with $k$ positive real
number. A kinetic growth process (KGT) of growing trails on the cubic
lattice maps to one temperature of the $k_G \simeq 4.15$ equilibrium
model. We find that the collapse is second-order if $k < k_G$ and
first-order if $k > k_G$. This resembles the triangular lattice
finding (although the nature of the second order transition is
different). Interestingly, the low temperature phase for both $k<k_G$ and
$k>k_G$ seems to be a disordered globular state. 

Exactly at $k = k_G$, the finite size scaling picture is
particularly intriguing: the energy distribution displays a double
peak form, indicating a first order type transition but we observe
different values for the {\em shift} exponent $\psi \simeq 1/2$ and
the \emph{width} ({\em crossover}) exponent $\phi \simeq 1$. The
thermodynamic limit location of this first order transition when $k =
k_G$ is the temperature $T_G$ that maps to the kinetic growth
process. However, if one simulates directly at the point $k=k_G$ and $T=T_G$ then the finite
size scaling encountered is entirely second-order like and shows no
sign of the first-order transition which dominates in the
thermodynamic limit. This can be understood by appreciating that the
finite size transition region shrinks quicker than its centre
approaches the limiting temperature.

These results help to illuminate previous contradictory work for interacting trails on the
diamond lattice \cite{prellberg1995b-:a,grassberger1996a-a}. As the coordination number of the diamond lattice equals 4, trails can
only interact through doubly visited sites. The collapse point of interacting trails on the
diamond lattice at $\omega=3$ was identified with the kinetic growth process. In \cite{prellberg1995b-:a} it
was shown that the scaling of the specific heat at the kinetic growth points for the diamond and simple
cubic lattices was indistinguishable. However, simulations of interacting trails on the diamond lattice
showed the emergence of a first-order phase transition \cite{grassberger1996a-a}. The scenario we describe here for KGT on the
simple cubic lattice clearly mimics these results. We are now able to understand the existence of both of these
behaviours through the breaking of crossover scaling. 

One last observation we can make is that interacting trails have been simulated in high dimensions \cite{prellberg2001a-:a,owczarek2003a-:a} and also
demonstrate the breakdown of crossover scaling. The behaviour in high dimensions has been shown to be consistent with
a self-consistent mean-field theory, which also displays bimodal energy distributions, though these do not lead to real
first-order transitions in high dimensions. While that theory cannot not be
applicable to $k=k_G$ in three dimensions (it predicts shift and width exponents both equal to $1/2$), it would be 
interesting to formulate a mean-field theory of the transition that occurs for our eISAT model when $k=k_G$. 
Consequently this may imply that the upper critical dimension for the $k=k_G$ eISAT models is less than three, and may in 
fact be two. Here we point out the numerical observation of confluent logarithms in the two-dimensional kinetic 
growth trails \cite{owczarek1995a-:a,owczarek2006c-:a}, lending further support to that assertion as logarithmic corrections typically appear at the
upper critical dimension of a phase transition.

\section{acknowledgment}

Financial support from the Australian Research Council
via its support for the Centre of Excellence for Mathematics
and Statistics of Complex Systems is gratefully acknowledged
by the authors. 
The simulations were performed on the
computational resources of the Victorian Partnership for Advanced
Computing.
A.L.O. thanks the School of
Mathematical Sciences, Queen Mary University of London
for hospitality.

\end{document}